\documentclass[preprintnumbers,amsmath,amssymb]{revtex4}
\usepackage{epsfig}
\usepackage{color}
\usepackage{dsfont}
\usepackage[colorlinks,urlcolor=blue,linkcolor=blue,citecolor=red]{hyperref}
\begin{document}
\setlength{\voffset}{1.0cm}
\title{Phase diagram of the massless Gross-Neveu model with two components}
\author{Michael Thies\footnote{michael.thies@gravity.fau.de}}
\affiliation{Institut f\"ur Theoretische Physik, Universit\"at Erlangen-N\"urnberg, D-91058, Erlangen, Germany}
\date{\today}

\begin{abstract}
Recently, Benini, Mamroud, Reis and Serone have presented exact results for the Gross-Neveu model at
finite density, both at finite $N$ and in the large $N$ limit. Generalizing previous studies, they introduce a chemical
potential acting only on a subset of fermion flavors --- the two-component Gross-Neveu model. Here, we 
take up this idea and extend the semiclassical study to finite temperature. The full (large $N$) phase diagram 
including inhomogeneous phases is constructed without solving the thermal Hartree-Fock problem.
All phase boundaries can be found on the basis of various kinds of stability analyses. As a result,
we find a qualitative change of the phase diagram at some critical filling fraction.
\end{abstract}

\maketitle

\section{Introduction} 
\label{sec1}

The Gross-Neveu (GN) model \cite{B1} is the quantum field theory of $N$ flavors of massless Dirac fermions in 1+1 dimensions, interacting via
a scalar-scalar four-fermion interaction,
\begin{equation}
{\cal L} =  \sum_{f=1}^{N} \bar{\psi}_f i \gamma^{\mu}\partial_{\mu} \psi_f + \frac{g^2}{2} \left( \sum_{f=1}^{N} \bar{\psi}_f \psi_f \right)^2.
\label{1.1}
\end{equation}
Its remarkable properties (renormalizability, dimensional transmutation, asymptotic freedom, discrete chiral symmetry breaking, bound states, integrability) have turned it into
an iconic toy model over the years. Notably the large $N$ limit has been explored thoroughly with semiclassical
methods, both at zero and finite temperature and density \cite{B2,B3,B4}. 

Recently, a lot of progress has been achieved in solving the GN model at finite $N$, using either lattice Monte Carlo methods \cite{B5,B6}, or by exploiting
the integrability of the model \cite{B7,B8,B9}. By and large, the semiclassical findings have been confirmed and limitations inherent in the large $N$ treatment
are by now better understood. An interesting generalization of the usual treatment of matter at finite density was proposed by Benini et al. \cite{B10,B11}. These authors
divide up the $N$ fermion flavors into two groups. The first $a$ fermions are called ``charged", the remaining $N-a$ ones ``neutral", and a chemical potential
is assigned to charged fermions only. In this way, important insights into renormalons and other features of the Bethe ansatz solution 
of the GN model with a finite number of flavors were obtained. For the present work, the implications for the semiclassical large $N$ treatment are most
relevant. The authors identified the ground state at finite density with a crystal composed of Dashen-Hasslacher-Neveu (DHN) baryons \cite{B2} with fermion number $a$.
Previously, the chemical potential had always been assumed to act on all $N$ flavors alike. In that case, the ground state crystal consists of baryons with maximal
fermion number $N$ (kinks and antikinks) only. Introducing two components thus enables one to generalize the notion of ``partially filled bound states" from the DHN baryon
to ``partially filled valence bands" of a periodic mean field in dense matter. 

That this concept is quite useful has already been confirmed in a preceding paper \cite{B12}. We have applied it to the zero temperature chiral GN model (with U(1)
chiral symmetry). It turned out that the twisted kink crystal of Basar and Dunne \cite{B13,B14} is the self-consistent mean field for fermion matter with filling fraction $\nu=a/N<1$. 
The fact that the low density limit of this potential looks like an array of partially filled DHN baryons had been known before, but how to realize this thermodynamically 
was unclear. Apparently, the twisted kink crystal is exactly the mean field it takes to describe a system with filling fraction $\nu<1$. For $\nu=1$, it reduces
to the simpler chiral spiral \cite{B15}. 

What happens if one heats up a system with a finite density of solitons? In the standard one-component GN model (\ref{1.1}) with a common chemical potential
for all flavors, this is well known. The phase diagram in the ($\mu,T$) plane features three phases, a massless one, a massive one and an
inhomogeneous one, separated by second order phase boundaries \cite{B16}. We shall recall it below in more detail. By contrast, nothing is known yet about 
the phase diagram for $\nu<1$, to the best of our knowledge. It is the purpose of the present paper to fill this gap.

Judging from the experience with the massless and massive GN models \cite{B17,B18}, the mean field at finite $T$ is likely to have the same functional form as at $T=0$.
There, the answer for the GN model is already known from Refs.~\cite{B10,B11}.  The mean field has the same analytical expression as that of the massive GN model.
It has long been known that partially filled DHN baryons of the {\em massless} GN model have the same shape as baryons
with maximal fermion number in the {\em massive} model \cite{B19,B20}. Apparently, a similar relationship holds for finite density systems. 

As is known from the massive GN model, a full Hartree-Fock (HF) calculation with such a mean field is elaborate \cite{B18}. It is necessary to minimize the grand canonical potential
with respect to three parameters at each point in 
($\mu,T,\gamma$) space, with $\gamma$ the confinement parameter related to the bare mass \cite{B18}. Here, we have to replace $\gamma$ by the filling fraction $\nu$, 
so that a full solution of the generalized GN model would be equally challenging. Fortunately, if we are only interested in the phase diagram and not in 
observables off the critical lines, efficient shortcuts are available. In the case of the GN model with a standard chemical potential, this was demonstrated only recently \cite{B21}. 
The full phase diagram of both the massless and the massive GN models have been reproduced without HF calculation and without
reference to the analytically known mean fields. The main idea is to use stability analyses with respect to different kinds of instabilities, 
both perturbative and non-perturbative ones. We shall explain these methods, some of  which have been explored in the context of GN models before \cite{B22,B23},
and apply them to the present generalization to arbitrary filling parameter. This will enable us to present the full phase diagram of the massless 
GN model for different values of $\nu$ with a modest effort.

The paper is organized as follows. In Sect.~\ref{sec2}, we solve the generalized problem under the (unjustified) assumption
of unbroken translational invariance. This is not only meant as a warm-up, but also serves to locate critical points
and phase boundaries possibly shared by the full calculation. We will learn that depending on $\nu$, either all
phase boundaries are connected, or they exhibit two disconnected branches. Accordingly, we divide the study of the full
phase diagram into two sections. Sect.~\ref{sec3} is devoted to the phase diagram of the full theory in the
connected region, Sec.~\ref{sec4} in the disconnected region of $\nu$-values. It is followed by a summary in Sec.~\ref{sec5}.
 
\section{Phase diagram assuming homogeneous condensates} 
\label{sec2}

Let us first construct the phase diagram of the two-component GN model assuming unbroken translation invariance. 
As is known from the one-component case, this may give clues about the full solution exhibiting inhomogeneous condensates.
The homogeneous phase diagram is likely to share with the full one certain phase boundaries and critical points.

The renormalized grand canonical potential density for the GN model with homogeneous mean field $M$ reads
\begin{equation}
N^{-1} \Psi_{\rm eff}^{(1)} (\beta,\mu)  =   \frac{M^2}{4\pi} \left( \ln M^2 -1\right) - \frac{1}{\beta \pi} \int_0^{\infty}dk \ln \left[ \left( 1+e^{-\beta(E_k-\mu)}\right) \left(1+e^{-\beta(E_k+\mu)} \right) \right].
\label{2.1}
\end{equation}
Here, $M$ plays the role of the dynamical fermion mass (set equal to 1 in the vacuum),  $E_k  =  \sqrt{k^2+M^2}$ is the single particle energy,
$\beta=1/T$ the inverse temperature, and $\mu$ the chemical potential for fermion number. The superscript (1) on $\Psi_{\rm eff}^{(1)}$ stands for one component,
i.e., a common chemical potential is introduced for all fermion flavors. The generalization to the two-component system where the chemical potential
acts only on a subset of $a=\nu N$ flavors is straightforward,
\begin{equation}
\Psi_{\rm eff}^{(2)}(\nu;\beta,\mu) = (1-\nu) \Psi_{\rm eff}^{(1)} (\beta,0)+ \nu  \Psi_{\rm eff}^{(1)} (\beta,\mu).
\label{2.2}
\end{equation}
Consider first the expected perturbative phase boundary between the chirally symmetric phase ($M=0$) and the chirally broken phase ($M>0$), defined by the condition
\begin{equation}
\frac{\partial}{\partial M^2} \left. \Psi_{\rm eff}^{(2)}(\nu;\beta,\mu)\right|_{M=0} = 0.
\label{2.3}
\end{equation}

\begin{figure}
\begin{center}
\epsfig{file=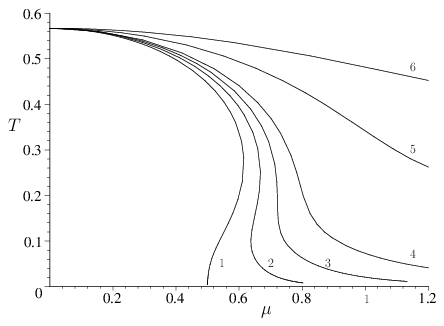,height=5cm,angle=0}
\caption{Perturbative phase boundaries of two-component GN model with homogeneous condensates, separating the massless from the massive phase.
Labels 1-6 correspond to filling fractions $\nu= 1.0, 0.90, 0.8271, 0.75, 0.50, 0.25$, respectively.}
\label{fig1}
\end{center}
\end{figure}

\begin{figure}
\begin{center}
\epsfig{file=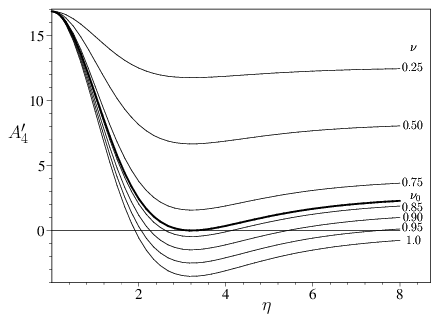,height=5cm,angle=0}
\caption{Determination of the tricritical point using the zero's of $A_4'$, Eq.~(\ref{2.7}), for a range of $\nu$-values. At $\nu=\nu_0 = 0.8271$ a bifurcation takes place and the 
tricritical point disappears.} 
\label{fig2}
\end{center}
\end{figure}

\begin{figure}
\begin{center}
\epsfig{file=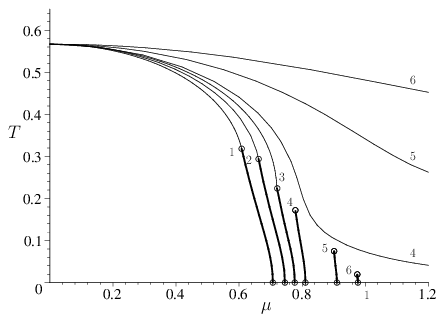,height=8cm,angle=0}
\caption{Full phase diagram of the two-component GN model with homogeneous condensates only. The labelling 1-6 is the same as in Fig.~\ref{fig1}.
Starting from the bifurcation curve 3 ($\nu=0.8271$), thin (second order) and fat (first order) curves get disconnected, the tricritical point being
replaced by a critical end point. The base points at $T=0$ are taken from \cite{B12}.} 
\label{fig3}
\end{center}
\end{figure}

Here, it is more convenient to turn to Ginzburg-Landau (GL) theory. For a uniform system, this is just a Taylor expansion in powers of $M$, sufficient to determine the exact perturbative
phase boundary. We recall the well-known GL expression for the (homogeneous) one-component system to the order needed here \cite{B17},
\begin{eqnarray}
N^{-1} \Psi_{\rm GL}^{(1)}(T,\mu) & = & \alpha_0(T,\mu)+  \alpha_2(T,\mu) M^2 + \alpha_4(T,\mu) M^4,
\nonumber \\
\alpha_0(T,\mu) & = & - \frac{\pi}{6} T^2- \frac{\mu^2}{2\pi}- \frac{1}{4\pi},
\nonumber \\
\alpha_2(T,\mu) & = & \frac{1}{2\pi}\left[ \ln (4\pi T)+ \Re \psi(z) \right],
\nonumber \\
\alpha_4(T,\mu) & = & - \frac{1}{2^6 \pi^3 T^2} \Re \psi(2,z), \quad z  =  \frac{1}{2} + \frac{i \eta}{2 \pi}, \quad \eta =  \frac{\mu}{T}.
\label{2.4}
\end{eqnarray}
It is given in terms of standard Digamma and Polygamma functions, $\psi(z)=\psi(1,z)$ and $\psi(n,z)$ with $n=2$.
This yields the following GL effective action for the two-component system, 
\begin{eqnarray}
\Psi_{\rm GL}^{(2)}(\nu;T,\mu) & = & (1-\nu) \Psi_{\rm GL}^{(1)}(T,0) + \nu \Psi_{\rm GL}^{(1)}(T,\mu)
\nonumber \\ 
& = & A_0(\nu;T,\mu)+  A_2(\nu;T,\mu) M^2 + A_4(\nu;T,\mu) M^4,
\nonumber \\
A_n(\nu;T,\mu) & = & (1-\nu) \alpha_n(T,0) + \nu \alpha_n(T,\mu) .
\label{2.5}
\end{eqnarray}
The condition for the perturbative phase boundary, Eq.~(\ref{2.3}), amounts to $A_2(\nu;T,\mu)=0$ or
\begin{equation}
0 = \ln (4 \pi T) + (1-\nu) \Psi(1/2)  + \nu \Re \Psi(z) , \quad \Psi(1/2)= -{\rm C} -2 \ln 2
\label{2.6}
\end{equation}
(C=Euler constant). Since $z$ depends only on $\eta=\mu/T$, Eq.~(\ref{2.6}) can be solved explicitly for $T(\eta)$. 
This is sufficient to plot $T$ vs. $\mu=T\eta$ with Maple using the parametric plot option, without need to solve 
transcendental equations. This step goes through for any $\nu$. The resulting dependence of the perturbative phase boundary
on $\nu$ is illustrated in Fig.~\ref{fig1}. With decreasing $\nu$ the curves get smoother, until they would reach a horizontal line
$T=T_c=e^{\bf C}/\pi$ at $\nu=0$ where nothing depends on $\mu$ anymore. From the standard treatment of the GN model ($\nu=1$), it is
known that the perturbative phase boundary ends at a tricritical point beyond which it goes over into a (non-perturbative)
first order phase boundary \cite{B24}. The position of a tricritical point can be found by
requiring $A_4=0$ in addition to $A_2=0$. Let us see what happens as we decrease $\nu$. The function $A_4$ reads
\begin{equation}
A_4 = \frac{(1-\nu) 14 \zeta(3) - \nu \Re \Psi(2,z)}{64 \pi^3 T^2} := \frac{A_4'(\nu,\eta)}{64 \pi^3 T^2}. 
\label{2.7}
\end{equation}
In Fig.~\ref{fig2} we have plotted $A_4'$ against $\eta$ for all values of $\nu$ used in the present work. 
Starting from $\nu=1$ and decreasing $\nu$, one first finds two solutions of the condition $A_4'=0$. The one at lower $\eta$
defines a tricritical point, the upper solution being unphysical. At a critical value  $\nu=\nu_0$, a bifurcation takes place ---  the two solutions coalesce
and there is no solution for $\nu<\nu_0$. As can be inferred  from Fig.~\ref{fig2}, the condition determining $\nu_0$ (in addition to $A_2=0,A_4=0$) 
is
\begin{equation}
\frac{\partial}{\partial \eta} A_4'(\nu,\eta) = 0
\label{2.8}
\end{equation}
or, equivalently, 
\begin{equation}
\Im \Psi(3,z) = 0.
\label{2.9}
\end{equation}
All three conditions together give the value $\nu_0$ of the bifurcation point and the coordinates ($\mu_0,T_0$) of
the tricritical point there. One finds
\begin{equation}
\nu_0=0.8271, \quad \mu_0 = 0.7202, \quad T_0=0.2240
\label{2.10}
\end{equation}
For $\nu<\nu_0$, there is no tricritical point anymore and the curves shown in Fig.~\ref{fig1} continue forever.

In the appendix of Ref.~\cite{B12} the same model has already been solved at $T=0$. A first order phase transition was found 
for any $\nu$ at a critical chemical potential $\mu_c$ depending on $\nu$. The dynamical mass changes discontinuously from the 
vacuum value to a lower value.  At $\nu=1$, the mass drops to 0 at $\mu_c=1/\sqrt{2}$, the base point of a first order phase boundary going
upwards in temperature and ending at the tricritical point. To study the corresponding curve for other values of $\nu$, one has to go back to the full expression for
the grand canonical potential, Eqs.~(\ref{2.1},\ref{2.2}), and locate points in the ($\mu,T$)-plane with two degenerate minima.
The result of such an analysis is as follows. For $1 \ge \nu \ge \nu_0$, the situation stays similar to $\nu=1$,
confirming the existence of a tricritical point. For lower $\nu$-values, the first order line does not reach the perturbative 
phase boundary but terminates at a critical point within the homogeneous massive phase. Thus, at the bifurcation value
$\nu_0$, the tricritical point moves away from the perturbative boundary, getting converted into a critical endpoint.
Some representative phase boundaries are shown in Fig.~\ref{fig3}. The base points are taken from Ref.~\cite{B12}.  
Tricritical points are evaluated as discussed above. The critical endpoints and first order curves are numerical results.
The coordinates of all endpoints of the first order curves determined in this work are listed in Table~\ref{tab1}.

\begin{center}
\begin{table}
\begin{tabular}{|c|c|c|c|}
\hline
$\nu$  & $\mu$ at $T=0$ &  $\mu_{\rm crit}$  & $T_{\rm crit}$  \\
\hline
1.0 & .7071 & .6082 & .3183 \\
.95 & .7255 & .6330 & .3083 \\
.90 & .7453 & .6620 & .2938 \\
.85 & .7665 & .6978 & .2675 \\
$\nu_0$ & .7765 & .7202 & .2240 \\
.75 & .8104 & .779 & .172 \\
.50 & .9102 & .902 & .074 \\
.25 & .9768 & .975 & .019 \\
\hline
\end{tabular}
\caption{End points of all first order lines computed in this work ($\nu_0=.8271$). The first order line shrinks to the point $\mu=1,T=0$)
in the limit $\nu \to 0$.}
\label{tab1}
\end{table}
\end{center}

Concluding this section, we have seen that the phase diagram of the homogeneous system features two qualitatively different regimes, depending 
on the value of $\nu$. In the region $1\ge \nu \ge \nu_0$, the first and second order phase boundaries are connected through a tricritical point.
In the region $\nu < \nu_0$, they are disconnected. The first order phase boundary rises from $T=0$ up to some critical point below the 
perturbative phase boundary. For simplicity, we refer to the two regions as ``connected" and "disconnected" from here on. It is not evident that  
this topological distinction remains valid once inhomogeneous condensates are allowed. Nevertheless, it is useful to discuss possible 
crystal phases separately in these two distinct regions. This is the topic of the following two sections.

\section{Full solution in the connected region}
\label{sec3}

\begin{figure}
\begin{center}
\epsfig{file=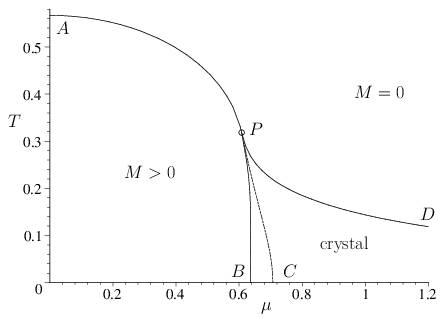,height=5cm,angle=0}
\caption{Phase diagram of the one-component GN model ($\nu=1$), adapted from \cite{B16}. The curve $PC$ from the first order line of the 
translationally invariant solution is replaced by two second order phase boundaries $PB$ and $PD$, whereas the second order line $AP$ 
is unchanged. The tricritical point $P$ is common to both phase diagrams.}
\label{fig4}
\end{center}
\end{figure}

To understand the GN phase diagram including inhomogeneous phases in the region $1 \ge \nu \ge \nu_0$, it is necessary to briefly recall the well-known case
$\nu=1$. Fig.~\ref{fig4} shows the phase diagram of the massless GN model with the standard chemical potential.
The curve $AP$ is the phase boundary between massless and massive homogeneous phases already known from the translationally
invariant calculation in Sec.~\ref{sec2}. Curve $PC$ is the first order phase boundary discussed above. It does not exist anymore in the full
phase diagram and has only been drawn to highlight the difference between homogeneous and full phase diagrams. 
Two new second order phase boundaries emerge, curve $PD$ separating the massless from the inhomogeneous phase and  curve
$PB$ separating the massive from the inhomogeneous phase. By abuse of language, we shall denote these two branches as ``horizontal" and ``vertical"
phase boundaries. The horizontal one can be determined by a standard perturbative stability analysis. The vertical one can be inferred from the
phase diagram of a single baryon by a different kind of stability analysis, based on the thermodynamics of a single baryon \cite{B12}.
Hence one does not need the full solution of the HF problem (available in this case) if one is only interested in mapping out the phase diagram,

Let us adapt the same procedures to the connected region $1 \ge \nu > \nu_0$. The curve $AP$ has already been constructed in Sec.~\ref{sec1}
and remains valid. The horizontal phase boundary $PD$ can be obtained as follows. For $\nu=1$, we follow \cite{B16} and use perturbation theory in the scalar potential,
using the ansatz
\begin{equation}
S(x) = 2 S_1\cos (2qx).
\label{3.1}
\end{equation}
For the one-component case ($\nu=1$),
the lowest order correction to the grand canonical potential density is
\begin{equation}
N^{-1} \delta \Psi^{(1)}(\beta,\mu) = \frac{S_1^2}{\pi}\left[ \ln(\Lambda) - {\cal P} \!\!\!\!\!\!\int_0^{\Lambda/2}dk \left( \frac{k}{k^2-q^2}\right) \frac{\sinh \beta k}{\cosh \beta k + \cosh \beta \mu}\right].
\label{3.2}
\end{equation}
The Cauchy principal value is a relic from almost degenerate perturbation theory (ADPT), necessary due to the gap in the spectrum of the periodic potential (\ref{3.1}).
For fixed $q$, requiring that (\ref{3.2}) vanishes
yields the curve along which the massless phase is unstable against spatial modulations with wave number $2q$.
The envelope of all these curves is the phase boundary $PD$. Rescaling of the variables according to
\begin{equation}
\kappa=\beta k, \quad \mu' = \beta \mu, \quad \Lambda'= \beta \Lambda, \quad q'=\beta q
\label{3.3}
\end{equation}
allows one to solve the condition that (\ref{3.2}) vanishes explicitly with the result
\begin{equation}
\beta_{\rm crit}(\mu',q') = \lim_{\Lambda' \to \infty} \Lambda' \exp \Phi^{(1)}(\mu',q',\Lambda')
\label{3.4}
\end{equation}
where
\begin{equation}
\Phi^{(1)}(\mu',q',\Lambda') = {\cal P}\!\!\!\!\!\!\int_0^{\Lambda'/2} d\kappa\left( \frac{\kappa}{(q')^2-\kappa^2}\right) \frac{\sinh \kappa}{\cosh \kappa + \cosh \mu'} .
\label{3.5}
\end{equation}
As in Sec.~\ref{sec2}, one can plot $T_{\rm crit}= 1/\beta_{\rm crit}(\mu',q')$ against $\mu_{\rm crit} = \mu'/\beta_{\rm crit}$ directly with Maple 
and draw a family of curves with varying $q$. The envelope of these curves can then easily be obtained numerically, leading to curve $PD$ in Fig.~\ref{fig4}. 
For the two-component GN model, we simply replace $\delta \Psi^{(1)}(\beta,\mu)$ in (\ref{3.2}) by
\begin{equation}
\delta \Psi^{(2)}(\nu; \beta,\mu) = (1-\nu) \delta\Psi^{(1)}(\beta,0) + \nu \delta\Psi^{(1)}(\beta,\mu)
\label{3.6}
\end{equation}
and $\Phi^{(1)}(\mu',q',\Lambda')$ in  (\ref{3.4}), (\ref{3.5}) by
\begin{equation}
\Phi^{(2)}(\nu; \mu',q',\Lambda')=(1-\nu)\Phi^{(1)}(0,q',\Lambda')+\nu\Phi^{(1)}(\mu',q',\Lambda').
\label{3.7}
\end{equation}
Everything else goes through as in the one-component case. 

The vertical phase boundary $PB$ is a different matter. Crossing this line in Fig.~\ref{fig4} from the left to the right, the homogeneous system
in the massive phase develops an instability towards forming a single DHN baryon with maximal fermion number (or, equivalently,
a kink). At $T=0$, this happens at $\mu$ equal to the corresponding baryon mass divided by $N$ ($\mu_c = 2/\pi$).
As explained in Ref.~\cite{B21}, the (second order) phase boundary $PB$ coincides with a first order phase boundary of a single baryon.
It has been determined in \cite{B21} for the one-component GN model as follows. Take as ansatz for the mean field (inspired by the $T=0$ DHN baryon)
\begin{equation}
S(x)=M\left(1+ y \left[\tanh (yMx-c_0)-\tanh(yMx+c_0) \right] \right), \quad c_0= \frac{1}{2} {\rm artanh} y.
\label{3.8}
\end{equation}
Here, $M$ is the thermal mass of the homogeneous solution, $y$ a variational parameter. 
The grand canonical potential of a single DHN baryon is then given by
\begin{eqnarray}
N^{-1} \Psi^{(1)}_B(\beta,\mu) & = & \frac{2yM}{\pi}\left( 1+ \ln \frac{m}{M}+\frac{\sqrt{1-y^2}}{y} \arctan \frac{\sqrt{1-y^2}}{y}\right) + \mu
 -  \frac{1}{\beta} \ln \left[ (1+e^{-\beta(E_0-\mu)})(1+e^{\beta(E_0+\mu)})   \right] 
\nonumber \\
&  &  + \frac{1}{\pi \beta} \int_0^{\infty} dk \frac{2yM}{k^2+y^2M^2}
\ln \left[    (1+e^{-\beta(E_k-\mu)})(1+e^{-\beta(E_k+\mu)})                  \right] ,
\label{3.9}
\end{eqnarray}
with $E_0=\sqrt{1-y^2}$ the energy of the discrete state. The parameter $y$ has to be determined by minimization. For the standard one-component Gross-Neveu model, 
one finds that $y$ jumps from 0 to 1 across a phase boundary interpolating nicely between the base point $M_B$ and the tricritical point 
of the homogeneous solution.

\begin{figure}
\begin{center}
\epsfig{file=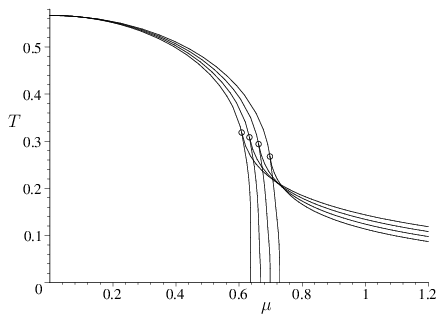,height=5cm,angle=0}
\caption{Generalizing the phase diagram from Fig.~\ref{fig4} to values of $\nu$ in the connected region $1 \le \nu \le \nu_0$. From left to right:
$\nu=1.0,0.95,0.90,0.85$}
\label{fig5}
\end{center}
\end{figure}

\begin{figure}
\begin{center}
\epsfig{file=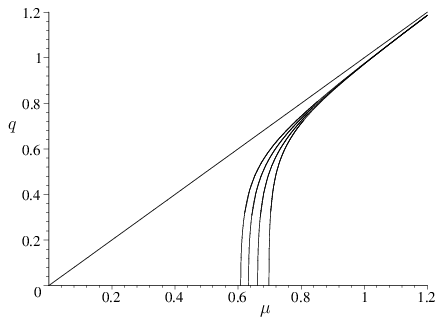,height=5cm,angle=0}
\caption{Wave number $q$ at which the massless phase is unstable. The result has been computed along the phase boundary, but is
plotted against the chemical potential. The $\nu$-values are the same as in Fig.~\ref{fig5}, again from left to right. The straight line is the 
asymptote $q=\mu$, the base points on the $\mu$-axis are the tricritical values.}  
\label{fig6}
\end{center}
\end{figure}

It is now easy to generalize this procedure to two components by setting
\begin{equation}
\Psi^{(2)}_B(\nu;\beta,\mu)= (1-\nu) \Psi^{(1)}_B(\beta,0) + \nu \Psi^{(1)}_B(\beta,\mu).
\label{3.10}
\end{equation}
From here on, everything follows the one-component case. For a given $T$ and $\nu$ one determines the chemical potential where the homogeneous
minimum (at $y=0$) and the second minimum at $1 \ge y >0$ are degenerate. This gives the critical chemical potential on the vertical curve $PB$.  
Here we do not have the full HF solution so that we cannot directly check the result. The fact that the vertical phase boundary 
obtained in this way ends precisely at the tricritical point of the homogeneous calculation with filling fraction $\nu$ is a strong indication
that the result is indeed correct. 

It is interesting to take the limit $T \to 0$ of (\ref{3.9}). In the region $\mu<1$ of interest here, one finds
\begin{eqnarray}
\lim_{\beta \to \infty} N^{-1}\Psi^{(2)}_B(\beta,\mu) & = &  \frac{2 y}{\pi}+ \frac{2 \sqrt{1-y^2}}{\pi} \arctan \frac{\sqrt{1-y^2}}{y} - \sqrt{1-y^2} 
\nonumber \\
& + & \nu \, \Theta(y-\sqrt{1-\mu^2}) (\sqrt{1-y^2}-\mu).
\label{3.11}
\end{eqnarray}
The minimum with respect to $y$ can be found by differentiation, yielding
\begin{equation}
\arctan \frac{\sqrt{1-y^2}}{y} = (1-\nu) \frac{\pi}{2} .
\label{3.12}
\end{equation}
This is precisely the relation between the parameter $y$ in the potential and the occupation fraction $\nu$ of the valence state
from the DHN baryon. The phase transition happens at $\mu=\mu_c$ where the right hand side of (\ref{3.11}) vanishes in
addition to (\ref{3.12}). Combining both conditions, we find 
\begin{equation}
\mu_c = \frac{2y}{\pi \nu} = \frac{M_B(\nu)}{\nu},
\label{3.13}
\end{equation}
as expected. Thus, at $T=0$, we see an instability against formation of a single DHN baryon with fermion number $\nu N$.

The result for several $\nu$-values between 1 and $\nu_0$ (where the tricritical point of the homogeneous calculation leaves
the perturbative phase boundary) is shown in Fig.~\ref{fig5} and shows no surprise. All the curves evolve in a gentle way
as long as one does not cross the bifurcation point $\nu_0$. The wave number of the critical fluctuation changes with $\mu$
along the phase boundaries $PD$. It vanishes at the tricritical point and approaches $q=\mu$ asymptotically.
This is illustrated in Fig,~\ref{fig6}, showing again a smooth variation between the standard
chemical potential ($\nu=1$) and the bifurcation point ($\nu=\nu_0$).

\section{Full solution in the disconnected region} 
\label{sec4}

\begin{figure}
\begin{center}
\epsfig{file=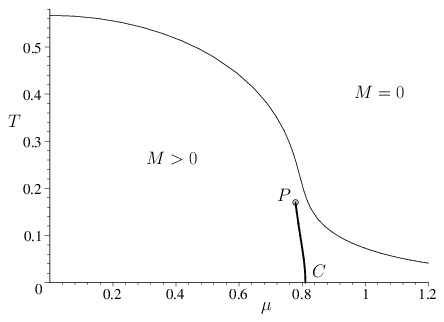,height=5cm,angle=0}
\caption{Homogeneous phase diagram of two-component GN model with $\nu=3/4$, same as curve 5 in Fig.~\ref{fig3}. The (thin) perturbative phase boundary
remains valid for all $\mu$, the (fat) first order phase boundary stays within the massive region, ending at the critical point $P$. The order parameter 
$M$ is discontinuous along curve $PC$.}
\label{fig7}
\end{center}
\end{figure}

If $\nu<\nu_0$, first and second order phase boundaries of the homogeneous phase diagram are disconnected. The tricritical point on the perturbative
phase boundary no longer exists. Instead, a critical point appears at the end of a first order phase boundary. We have redrawn the
homogeneous phase diagram at $\nu=3/4$ in Fig.~\ref{fig7}. How does this phase diagram change if we allow for inhomogeneous condensates?
We expect that the perturbative phase boundary remains valid. The critical point is now immersed into the massive homogeneous phase.
This situation is reminiscent of the massive GN model, except that the massless phase does not exist there. In the massive
GN model, it was found that the first order line disappears if one allows for crystal phases. It is replaced by two second order lines starting from the same critical point ($\mu_c,T_c$)
in a cusp and enclosing a crystal region. These lines are analogous to the horizontal and vertical branches in the connected region ($1 \ge \nu \ge \nu_0$).
Let us once again try to construct these curves using appropriate stability analyses. 

Consider first the horizontal curve. In the preceding section, the corresponding phase boundary signals instability of the massless phase
against crystallization. Here, instead. we are dealing with the instability of the massive phase, at least close to the point $P$. This can again be
dealt with by a perturbative stability analysis with minor changes due to the non-zero mass on the homogeneous side. Actually, the necessary formulas can be taken
over from previous work on the massive GN model. The relevant formulas can be found in Ref.~\cite{B25} on the massive chiral GN model (with U(1) chiral symmetry).
We take the condition for the instability, Eq.~(37) of Ref.~\cite{B25}, and set the pseudoscalar potential $P_1$ and the confinement parameter $\gamma$ equal to 0 
with the result
\begin{equation}
N^{-1} \delta \Psi^{(1)}(\beta,\mu)  =  \frac{E_q^2 S_1^2}{\pi} {\cal P}\!\!\!\!\!\! \int_0^{\infty} dk \frac{1}{E_k(k^2-q^2)}\left( \frac{1}{e^{\beta(E_k-\mu)}+1} + \frac{1}{e^{\beta(E_k+\mu)}+1}\right)
- \frac{E_q S_1^2}{2 \pi q} \ln \left( \frac{E_q-q}{E_q+q}\right).
\label{4.1}
\end{equation}
The step from here to the two-component system is once again trivial,
\begin{equation}
\delta \Psi^{(2)}(\nu;\beta, \mu) = (1-\nu) \Psi^{(1)}(\beta,0) + \nu \Psi^{(1)}(\beta, \mu).
\label{4.2}
\end{equation}
Expression (\ref{4.2}) and its derivative with respect to $q$ must vanish along the phase boundary.
Here one cannot solve the equation for $T$ explicitly like in the massless case, Sec.~\ref{sec3}, but has to find the solution numerically.

Now turn to the vertical phase boundary. From the massive GN model and Ref.~\cite{B21}, we also know that it is closely related to a first order line in the thermodynamics of a single baryon.
It should connect the critical point from the homogeneous calculation with the point  
\begin{equation}
\mu_c= \frac{2 \sin (\pi \nu/2)}{\pi \nu}
\label{4.3}
\end{equation}
at $T=0$, i.e., the mass of a DHN baryon (fermion number $\nu N$) divided by $\nu$. During the calculation, one has to minimize (\ref{3.9},\ref{3.10}) with respect to the variational parameter $y$.

\begin{figure}
\begin{center}
\epsfig{file=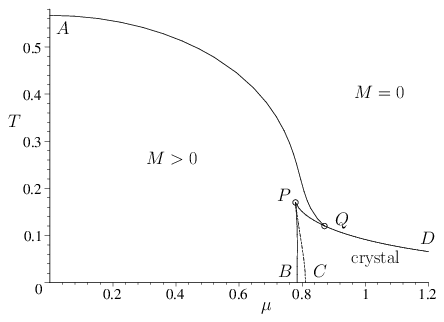,height=5cm,angle=0}
\caption{Full phase diagram of two-component GN model with $\nu=3/4$ including inhomogeneous region. The first order line $PC$ of Fig.~\ref{fig7}
is replaced by the second order line $PB$ (instability against single baryon formation). Line $PQ$: perturbative instability of massive phase against
periodic modulation. Curve $QD$: similar instability of massless phase. The curve $AQ$ and the point $P$ are common with Fig.~\ref{fig7}, point $Q$
is a new bifurcation point where two critical lines intersect.}
\label{fig8}
\end{center}
\end{figure}

In Fig.~\ref{fig8}, we show the result of such a calculation at $\nu=3/4$. We find indeed the two second order lines starting at the critical point in a cusp-like shape and 
enclosing a crystal phase, familiar from the massive GN model. However, there is one novel feature. The horizontal boundary intersects the perturbative homogeneous boundary
in the point $Q$, giving rise to a new kind of bifurcation as a function of $\mu$. Unlike $P$, the critical point $Q$ does not exist in the homogeneous 
phase diagram. The curve segment $PQ$ has been generated from the stability analysis of the massive phase and uses the 
thermal mass from the homogeneous phase diagram. The part $QD$ of the curve is the instability of the massless phase, as discussed
in the connected region. To see how things evolve with decreasing $\nu$, we also tried to find the bifurcation point $Q$ for other values of $\nu$.
It turns out that the distance between $P$ and $Q$ increases very rapidly with decreasing $\nu$, so that we could find the point $Q$ only down to $\nu \approx 0.70$
with our calculational method.

\begin{figure}
\begin{center}
\epsfig{file=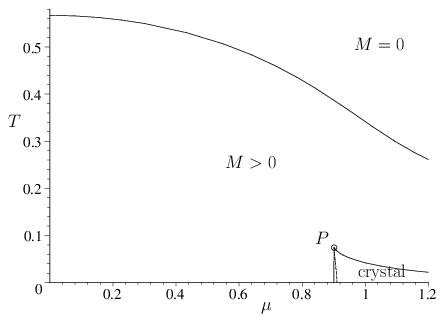,height=5cm,angle=0}
\caption{Full phase diagram of two-component GN model with $\nu=1/2$, showing that the crystal region shrinks rapidly with decreasing $\nu$.
The corresponding homogeneous phase boundary corresponds to curves 5 in Fig.~\ref{fig3}. The point analogue to $Q$ in Fig.~\ref{fig8} cannot be 
seen in the region explored here.}
\label{fig9}
\end{center}
\end{figure}

\begin{figure}
\begin{center}
\epsfig{file=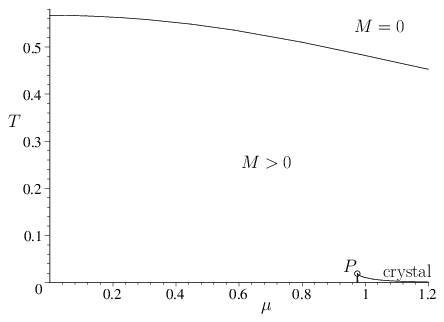,height=5cm,angle=0}
\caption{Like Fig.~\ref{fig9}, but for $\nu=1/4$. The crystal region is now tiny as could have been anticipated from the short first order line
in curve 6 of Fig.~\ref{fig3}.}
\label{fig10}
\end{center}
\end{figure}

If we proceed to smaller $\nu$-values, the inhomogeneous region shrinks rapidly, as shown in Figs.~\ref{fig9} and \ref{fig10}. The similarity of
these plots with those of the massive GN model is striking. A small occupation fraction of the valence levels quenches the crystal region in the
phase diagram in a similar way as a large confinement parameter (or bare fermion mass) does. We cannot rule out that an intersection
point $Q$ is present here as well, but it should move to very high chemical potentials where the width of the inhomogeneous island becomes
tiny.

Concluding this section, we remark that the classification into connected and disconnected phase diagrams from the homogeneous calculation
is not valid anymore, once one allows for inhomogeneous condensates. At least in the vicinity of the bifurcation point $\nu_0$,
we have seen that the phase diagram is again connected, see Fig.~\ref{fig8}.

\section{Summary} 
\label{sec5}

In this paper, we have taken up a recent proposal by Benini et al. \cite{B10,B11} of generalizing the (large $N$) GN model at finite density. The
idea is to leave the Lagrangian of the GN model unchanged, but impose a chemical potential that acts on a fraction $\nu$ of the $N$ fermion
flavors only. We refer to this extension as two-component GN model. It generalizes the notion of partially filled valence states familiar from the DHN baryon
to partially filled bands for a periodic array of such baryons. Results are already available at $T=0$, both for the GN model ($Z_2$ chiral symmetry \cite{B10,B11}) 
and the chiral GN model ($U(1)$ chiral symmetry \cite{B12}). The aim of the present work was to study the two-component GN model at finite temperature and
chemical potential in the large $N$ limit. Rather than performing a full thermal HF calculation, we have constructed the phase diagram exclusively with the help of  
different kinds of stability analyses. As is known from a recent study of the massless and massive one-component GN models, this is sufficient to
yield the exact phase boundaries and critical points, both perturbative and non-perturbative.

In a first step, we had to generalize the thermodynamic calculation assuming a homogeneous condensate. This simple exercise already revealed
a qualitatively new phase diagram as compared to the corresponding one-component calculation. As a function of $\nu$, a bifurcation
was identified at $\nu=\nu_0=0.8271$. For larger occupation fraction, the phase diagram looks similar to the one at $\nu=1$. A second order
phase boundary separating massless and massive phases ends at a tricritical point, where it goes over into a first order phase boundary with
a discontinuous jump in the fermion mass. For $\nu<\nu_0$, the tricritical point disappears and the perturbative
second order line does not end anymore. A first order line appears inside the massive region, terminating at a critical point. For $\nu=\nu_0$, this 
critical point touches the perturbative phase boundary and is converted into a tricritical point. The length of the first order line decreases
rapidly with decreasing $\nu$.

If one allows for inhomogeneous condensates, the distinction between the two regimes separated by the bifurcation at $\nu_0$ stays clearly
visible. For $\nu>\nu_0$, the stability analyses give a phase diagram looking very much like that of the one-component model. The perturbative
phase boundary between massless and massive homogeneous phases remains valid. The tricritical point of the homogeneous phase diagram also
stays in place. The first order line disappears, being replaced by a cusp-like structure of two second order lines, framing
a crystal region. In the homogeneous phase diagram, for $\nu<\nu_0$, a first order line ends inside the massive region at a critical point. This is reminiscent of the
massive (one-component) GN model, except that the massless region is absent there. The effect of inhomogeneous condensates is also similar
to what happens in the massive GN model. The first order line disappears and gets replaced by a pair of second order lines emanating from the critical point of the
homogeneous model and enclosing a crystal region. The size of this region shrinks rapidly with decreasing $\nu$. An interesting new effect
was found in the vicinity of the bifurcation, here at $\nu=3/4$. The horizontal branch of the second order lines intersects the
perturbative phase boundary in a point $Q$ where another type of bifurcation takes place, now as function of $\mu$ rather than $\nu$.
For the stability analysis, this means that one has to apply a perturbative stability analysis both to the massive phase and the massless phase.
Techniques formerly developed for massless and massive GN models have been used accordingly. Remarkably, in Fig.~\ref{fig8}, one finds four 
distinct second order phase boundaries requiring all four kinds of stability analysis known from the GN model: massless to massive ($AQ$),
massless to periodic ($QD$), massive to periodic ($PQ$) and massive to single soliton ($PB$). This underlines how much the extra freedom due to the parameter $\nu$ 
enriches the study of the phase diagram of the GN model.

\end{document}